\documentclass[prl,twocolumn]{revtex4}
\usepackage{amsfonts}
\usepackage{amsmath}
\usepackage{amssymb}
\usepackage{graphicx}

\begin{document}

\preprint{}
\title{Kinetic limitations of cooperativity based drug delivery systems}
\author{Nicholas A. Licata and Alexei V. Tkachenko}
\affiliation{Department of Physics and Michigan Center for Theoretical Physics,
University of Michigan, 450 Church Street, Ann Arbor, Michigan 48109}

\begin{abstract}
We study theoretically a novel drug delivery system that utilizes the
overexpression of certain proteins in cancerous cells for cell specific
chemotherapy. \ The system consists of dendrimers conjugated with "keys"
(ex: folic acid) which "key-lock" bind to particular cell membrane proteins
(ex: folate receptor). \ The increased concentration of "locks" on the
surface leads to a longer residence time for the dendrimer and greater
incorporation into the cell. \ Cooperative binding of the nanocomplexes
leads to an enhancement of cell specificity. \ However, both our theory and
detailed analysis of in-vitro experiments indicate that the degree of
cooperativity is \textit{kinetically limited}. \ We demonstrate that
cooperativity and hence the specificity to particular cell type can be
increased by making the strength of individual bonds \textit{weaker}, and
suggest a particular implementation of this idea. \ The implications of the
work for optimizing the design of drug delivery vehicles are discussed. \ 
\end{abstract}

\maketitle

Nanoparticle based drug delivery systems have attracted substantial
attention for their potential applications in cancer treatment (\cite%
{target1},\cite{target2},\cite{target3},\cite{target4},\cite{nanocluster}).
\ It is hoped that by selectively targeting cancer cells with
chemotherapeutic agents one can reduce side effects and improve treatment
outcomes relative to other drug delivery systems which do not discriminate
between normal and cancerous cells. \ For example, many epithelial cancer
cells are known to overexpress the folate receptor(\cite{ovarian},\cite%
{folate1},\cite{folate2},\cite{folate3},\cite{folate4}). \ A nanoparticle
with many folic acid ligands will preferentially bind to cancerous cells. \
A recent study\cite{pamam} of a potential drug delivery platform consisting
of generation 5 PAMAM\ dendrimers with different numbers of folic acid found
that multivalent interactions have a pronounced effect on the dissociation
constant $K_{D}$. \ This enhancement is the signature for cooperativity of
the binding, which should lead to a greater specificity to cancerous cells
in vivo. \ 

\begin{figure}[tbp]
\includegraphics[width=2.6676in,height=2.0104in]{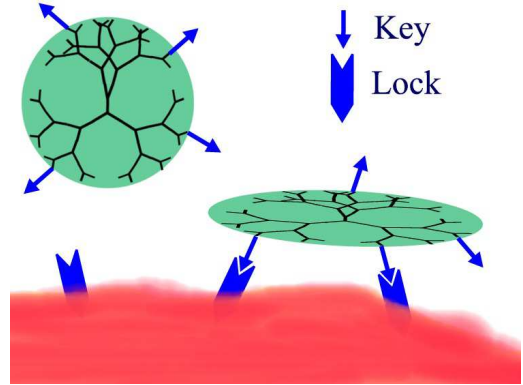}
\caption{(Color online). A picture of the dendrimer "key-lock" binding to
the cell membrane surface. \ }
\label{dendrimernew}
\end{figure}

In this letter we present a theoretical study of these key-locking
nanodevices (see Fig. \ref{dendrimernew}). \ We introduce the idea that
there are kinetic limitations to cooperativity-based drug delivery systems.
\ In vivo the finite timescale for endocytosis prevents arbitrarily high
cooperativity in the drug delivery system. \ In the first part we provide a
detailed analysis of the in vitro experiments\cite{pamam}. \ Although
enhancement of the association is the signature of greater cooperativity, in
this case it is due mostly to non-specific binding of the dendrimers to the
surface. \ Due to the finite time window of the experiments, only indirect
support can be offered to the notion of enhanced cooperativity. \ In the
second part we expand the notion of kinetically limited cooperativity to the
system in vivo. \ The equilibrium coverage of nanodevices on the cells is
related to the concentration of folate-binding proteins and the strength of
the key-lock binding. \ We quantify the preferential adsorption of
nanodevices to the cancerous cells, and discuss how kinetic effects prohibit
arbitrarily high cooperativity in the drug delivery system. \ The
implications of the work for designing new drug delivery vehicles with
enhanced specificity to cancerous cells are discussed. \ 

We now consider a simple model of the nanodevice system. \ A dendrimer with
a maximum of $M$ keys (e.g. folic acids) interacts with locks (e.g.
folate-binding proteins) in the cell membrane surface. \ A simple order of
magnitude estimate for $M\simeq 30$ can be obtained from the ratio of the
surface area of the dendrimer to the surface area of the folic acid. \ In
this way we implicitly take into account the excluded volume effect between
the keys. \ The free energy for the dendrimer connected to the surface by $m$
key-lock bridges is\cite{statmech} $\ $%
\begin{equation}
F_{m}=-k_{B}Tm\Delta \text{.}
\end{equation}%
The dimensionless energy parameter $\Delta $ contains information about the
binding energy of a single key-lock pair, and the entropy loss associated
with localizing a dendrimer on the cell-membrane surface. \ An estimate of $%
\Delta \simeq 17.5$ can be obtained from the dissociation constant of free
folic acid $K_{D}^{(o)}$ using the equilibrium relation between the
dissociation constant and the free energy change for the formation of a
single key-lock bridge, $K_{D}^{(o)}=\frac{1}{\xi ^{3}}\exp (-\Delta )$. \
Here $\xi ^{3}$ is the localization volume of an "unbound" key. \ Below we
determine the value $\xi \simeq 0.2nm$ from analysis of the in vitro
experiments, which was used to determine $\Delta $. \ 

\begin{figure}[tbp]
\includegraphics[width=3.5206in,height=2.6498in]{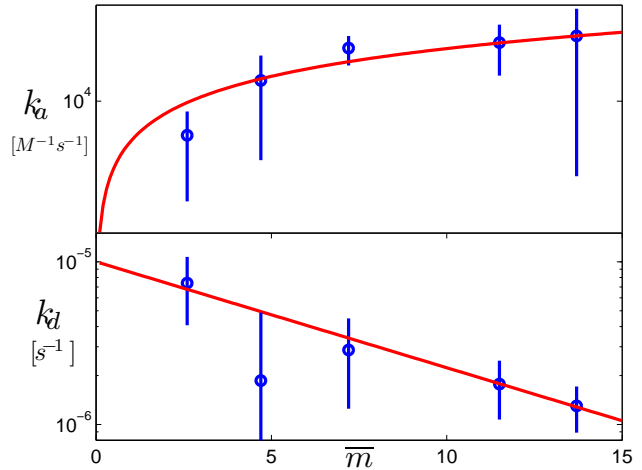}
\caption{(Color online). \ Top: Plot of the association rate constant (Eq. 
\protect\ref{kaeqn}) $k_{a}[M^{-1}s^{-1}]$ versus $\overline{m}$. \ Bottom:
Plot of the effective dissociation rate constant (Eq. \protect\ref{kdeffeqn}%
) $k_{d}[s^{-1}]$ versus $\overline{m}$. \ In the fit $%
k_{d}^{(o)}=10^{-5}[s^{-1}]$ and $\protect\alpha =0.15$. \ The experimental
data points are taken from Figure 5 in \protect\cite{pamam}.}
\label{kplot5}
\end{figure}

The measured association rate constant $k_{a}$ of the dendrimer with folic
acid is a factor of $10^{3}$ times greater than $k_{a}^{(o)}$ of free folic
acid. \ Only a factor of $\overline{m}$ can be attributed to the dendrimer
having many folic acids attached to it. \ Here $\overline{m}$ is the average
number of keys attached to the dendrimer. \ This pronounced enhancement of $%
k_{a}$ is the primary evidence for non-specific attraction between the
dendrimer and the surface. \ 
\begin{equation}
k_{a}=\overline{m}k_{a}^{(o)}\exp \left( \frac{-\epsilon _{0}}{k_{B}T}\right)
\label{kaeqn}
\end{equation}%
The non specific attraction $\epsilon _{0}$ accounts for the Van der Waals
attraction to the surface and hydrophobic enhancement. \ The experimentally
measured $k_{a}$ values are reproduced by a reasonable energy scale $%
-\epsilon _{0}\simeq 7k_{B}T$ (see Fig. \ref{kplot5}). \ 

We provide a simple explanation for the experimentally observed dependence
of the dissociation rate constant $k_{d}$ on $\overline{m}$. \ The
dissociation rate constant of free folic acid $k_{d}^{(o)}\sim 10^{-5}\left[
s^{-1}\right] $ provides a characteristic departure time of $%
1/k_{d}^{(o)}\simeq 30$ $hours$ for those dendrimers attached by a single
key-lock bridge. \ Moreover, the departure time for multiple bridge states
increases exponentially in $\Delta $, for two bridges it is $\exp (\Delta
)/k_{d}^{(o)}\simeq 10^{9}$ $hours$. \ Strictly speaking the relaxation is
multiexponential, with time constants for each bridge number. \ However, the
experimental $k_{d}$ values are well fit by a single exponential. \ On the
timescale of the experiment, we will only see the departure of dendrimers
attached by a single bridge. \ 

The experiment measures the departure rate of dendrimers which are connected
to the surface by a single bridge, but are unable to form an additional
connection. \ Consider a dendrimer attached to the surface by one key-lock
bridge. \ If the dendrimer has a total of $j$ keys, the probability that
none of the remaining $j-1$ keys can form bridges is $(1-\alpha )^{j-1}$. \
We now compute the probability $\alpha $ that a remaining key is available
to form a bridge. \ In the vicinity of the surface the dendrimer is a
disclike structure\cite{mecke} with radius $a\simeq 4.8nm$. \ By rotation of
the dendrimer about the first bridge, a key located at position $\rho $
searches the annulus of area $2\pi \rho \xi $ to find a lock. \ The
probability of encountering a lock in this region is $2\pi \rho \xi \sigma
_{o}$, where the surface density of the locks $\sigma _{o}\simeq \frac{16}{%
100nm^{2}}$. \ By averaging over the key location we obtain the final result%
\begin{equation}
\alpha =\frac{1}{a}\int_{0}^{a}2\pi \rho \xi \sigma _{o}d\rho \simeq \xi
a\sigma _{o}.  \label{alpha}
\end{equation}%
Assuming that during dendrimer preparation the attachment of folic acid to
the dendrimer is a Poisson process, the probability of a dendrimer having
exactly $j$ keys is $P_{j}(\overline{m})=\exp (-\overline{m})\overline{m}%
^{j}/j!$. \ The final result is obtained by averaging the probability that
no additional bridges can form over this distribution. \ The factor of $j$
counts the number of ways to make the first connection. \ 
\begin{equation}
k_{d}=k_{d}^{(o)}\frac{\sum\limits_{j=1}^{\infty }(1-\alpha )^{j-1}jP_{j}(%
\overline{m})}{\sum\limits_{j=1}^{\infty }jP_{j}(\overline{m})}%
=k_{d}^{(o)}\exp (-\alpha \overline{m})  \label{kdeffeqn}
\end{equation}

The formula predicts an exponential decay of the effective dissociation rate
constant with the average number of folic acids on the dendrimer, which
allows for a quantitative comparison to the experiment (see Fig. \ref{kplot5}%
). \ Using $\alpha \simeq 0.15$, we can determine the localization length $%
\xi \simeq 0.2nm$ for locks in the experiment from Eq. \ref{alpha}. \ This
estimate for $\xi $ is physically reasonable, and comparable to the bond
length of the terminal group on the dendrimer. \ 

Similar to the finite timescale of the experiments in vitro\cite{pamam}, in
vivo the endocytosis time provides kinetic limitations to cooperative
binding. \ In equilibrium the concentration of dendrimers on the cell
surface $n$ is related to the concentration of dendrimers in solution $%
c_{sol}$ through the association constant $K_{A}=n/(\sigma _{o}c_{sol})$. \
Although it is tempting to use our in vitro results to define the
association constant as $K_{A}=k_{a}/k_{d}$, this approach is only valid
provided there is a single rate for both association and dissociation. \
Because the dendrimer can form multiple bridges, there are many different
rate constants. \ We present a partition function method which accounts for
the multiple rate constants in the problem, and for the possibility that in
vivo there is surface diffusion of locks. \ 

To proceed we construct a vector $\mathbf{s}$ of length $M$, which is a list
of the possible sites folic acid can attach to the dendrimer. \ If a folic
acid is present at site $i$ we have $s_{i}=1$, and otherwise $s_{i}=0$. \
The concentration of dendrimers on the cell surface $n$ is proportional to
the partition function of the system. \ 
\begin{gather}
n=\frac{c_{sol}\xi ^{3}}{A}\sum\limits_{m=1}^{\infty }\int \frac{d^{2}%
\mathbf{r}_{1}\cdots d^{2}\mathbf{r}_{m}}{m!}\sum_{i\neq j\neq \cdots \neq
p}s_{i}\cdots s_{p} \\
\times \sigma (\mathbf{r}_{1})\cdots \sigma (\mathbf{r}_{m})\exp \left[
m\Delta -\frac{\epsilon _{0}+\varepsilon _{ij\cdots p}(\mathbf{r}_{1},\cdots
,\mathbf{r}_{m})}{k_{B}T}\right]  \notag
\end{gather}%
Here $\sigma (\mathbf{r})$ is the surface density of locks on the cell
membrane at position $\mathbf{r}$, and $A$ denotes the total area of the
cell membrane. \ The energy $\varepsilon _{ij\cdots p}(\mathbf{r}_{1},\cdots
,\mathbf{r}_{m})$ that appears in the Boltzmann weight is the elastic energy
penalty required to form multiple bridges. \ The point is that in solution
the dendrimer is roughly spherical, but must flatten to a pancake like shape
to form multiple connections with the cell surface\cite{mecke}. \ 

The ensemble averaging is performed by assuming that during nanodevice
preparation the attachment of folic acid to the dendrimer is a Poisson
process. $\ $In this case $\left\langle s_{i}\right\rangle =\frac{\overline{m%
}}{M}$ is given by the success probability that a folic acid attaches to the
dendrimer, and the$\ m$ point correlator $\left\langle s_{i}s_{j}\cdots
s_{p}\right\rangle =\left( \frac{\overline{m}}{M}\right) ^{m}$. \ In other
words, the probability of attachment of a given folic acid to a terminal
group on the dendrimer is unaffected by the presence of other folic acids up
to an exclusion rule which has already been taken into account. \ If the
interaction potential between locks in the cell membrane is $V(\mathbf{r}%
_{1,}\cdots ,\mathbf{r}_{m})$ we have $\left\langle \sigma (\mathbf{r}%
_{1})\cdots \sigma (\mathbf{r}_{m})\right\rangle =\left( \sigma _{o}\right)
^{m}\exp \left[ -V(\mathbf{r}_{1,}\cdots ,\mathbf{r}_{m})/k_{B}T\right] $. \
By performing the ensemble averaging we arrive at the result for the
equilibrium coverage $n_{m}^{eq}$ of dendrimers connected to the cell
surface by $m$ bridges. \ 
\begin{align}
n_{m}^{eq}& =\frac{c_{sol}\xi ^{3}}{m!}\left( \frac{\overline{m}\sigma _{o}}{%
M}\right) ^{m}\exp \left[ m\Delta -\frac{\epsilon _{0}}{k_{B}T}\right]
\sum_{i\neq j\neq \cdots \neq p}  \notag \\
& \times \int d^{2}\mathbf{r}_{2}\cdots d^{2}\mathbf{r}_{m}\exp \left[ -%
\frac{\left( \varepsilon _{ij\cdots p}+V\right) \left( \mathbf{0},\mathbf{r}%
_{2},\cdots ,\mathbf{r}_{m}\right) }{k_{B}T}\right]   \notag \\
n_{m}^{eq}& =\frac{c_{sol}\xi }{m!}\left( \frac{\overline{m}%
K_{A}^{(o)}\sigma _{o}}{\xi }\right) ^{m}\exp \left[ \frac{-\left( \epsilon
_{el}^{(m)}+\epsilon _{0}\right) }{k_{B}T}\right] 
\end{align}%
\ \ \ \ \ \ Here $K_{A}^{(o)}=1/K_{D}^{(o)}$ $=\xi ^{3}\exp (\Delta )$ is
the association constant of free folic acid which has been measured
experimentally. \ Defined in this manner, $\exp (-\epsilon
_{el}^{(m)}/k_{B}T)$ has a physical interpretation as the Boltzmann weight
for the elastic energy of the optimal $m$ bridge configuration. \ The
membrane surface can only accomodate a finite number of locks in the
vicinity where the dendrimer is attached\cite{pamam}. \ As a result $%
n_{m}^{eq}=0$ for $m>m_{\max }$ since forming additional key-lock pairs
would require deforming the dendrimer into configurations prohibited by
elastic stress and steric hindrance. \ The calculation of the equilibrium
coverage above is applicable with and without diffusion of locks in the cell
membrane. \ In the regime of fast diffusion the locks are free to
diffusively explore the surface. \ Their positions are ergodic variables,
and the overall ensemble averaged equilibrium coverage counts the Boltzmann
weights for different lock configurations. \ In the regime of slow
diffusion, locks are immobilized in the cell membrane. \ This is the
relevant situation when the locks have phase separated into protein rich
(lipid rafts) and protein poor phases. \ 

When kinetic effects are taken into account, these two regimes are
drastically different. \ When locks are diffusing, the dendrimer is able to
attain the maximum cooperativity $m_{\max }$. \ After the dendrimer makes
the first connection, it simply waits for locks to diffuse in the vicinity
of available keys to make additional connections. \ In the absence of
diffusion, the optimal configuration can only be obtained by multiple
binding and unbinding events, the timescale for which is prohibitively long.
\ This is the case for lipid rafts where the locks are immobilized similar
to the experiments in vitro\cite{pamam}, and the dendrimer is unable to
attain the maximum cooperativity. \ This is the kinetic origin of limited
cooperativity in the drug delivery system. \ 

We now quantify the preferential attachment of nanodevices to the cancerous
cells, taking into account kinetic effects. \ Let $n_{m}$ denote the
concentration of dendrimers attached to the cell by $m$ bridges. \ We can
construct a differential equation for $n_{m}$ by considering linear response
to the deviation from thermal equilibrium $n_{m}^{eq}$. \ 
\begin{equation}
\frac{dn_{m}}{dt}=k_{d}^{(m)}(n_{m}^{eq}-n_{m})-\gamma n_{m}
\end{equation}%
Here $\gamma $ is the rate for endocytosis\cite{endocytosis}. \ The
dissociation rate constant $k_{d}^{(m)}$ for breaking all $m$ bridges is: \ 
\begin{equation}
k_{d}^{(m)}=m\frac{k_{a}^{(o)}}{\xi ^{3}}\exp \left( \frac{\epsilon
_{el}^{(m)}}{k_{B}T}\right) \exp (-m\Delta )
\end{equation}%
The steady state concentration $n_{m}^{ss}$ is the solution to $\frac{dn_{m}%
}{dt}=0$. \ As a result we obtain the total coverage $n$ of dendrimers on
the cell surface in the following form:%
\begin{equation}
n=\sum_{m=1}^{m_{\max }}n_{m}^{ss}=\sum_{m=1}^{m_{\max }}\frac{n_{m}^{eq}}{%
1+\gamma /k_{d}^{(m)}}
\end{equation}

\begin{figure}[tbp]
\includegraphics[width=2.9966in,height=2.2563in]{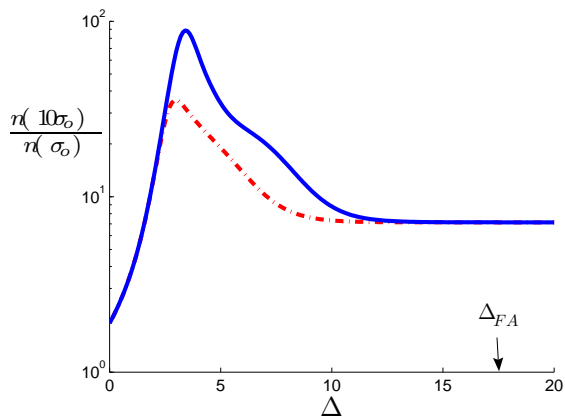}
\caption{(Color online). \ The ratio of surface concentrations of dendrimers
on cancerous to normal cells $\frac{n(10\protect\sigma _{o})}{n(\protect%
\sigma _{o})}$ as a function of $\Delta $ with $r=10$. \ The dotted line
corresponds to an endocytosis time $1/\protect\gamma =1\left[ hr\right] $
and the solid line is $1/\protect\gamma =10\left[ hr\right] $. \ Here $%
\overline{m}=15$, $m_{\max }=4$, $\protect\xi =3[nm]$, and $\protect\sigma %
_{o}=2\times 10^{-3}[nm^{-2}]$. \ $\protect\varepsilon _{el}^{(m)}=3k_{B}T$
for $m\geq 3$ bridges. \ }
\label{logratio}
\end{figure}

We now have a means to discuss the preferential attachment of dendrimers to
the cancerous cell. \ The folate binding proteins on the cancerous cell are
overexpressed, i.e. if their concentration on the normal cell is $\sigma
_{o} $, their concentration on the cancerous cell is $r\sigma _{o}$ with $%
r>1 $. \ The value of $r$ is determined by the biology, and cannot be
changed by the experimenter. \ To quantify the preferential binding of the
dendrimer to the cancerous cells we calculate the ratio of coverage on
cancerous to normal cells $\frac{n(r\sigma _{o})}{n(\sigma _{o})}$. \ Values
of this ratio greater than $r$ indicate the nature of cooperative dendrimer
binding (see Fig. \ref{logratio}).

\begin{figure}[tbp]
\includegraphics[width=2.6676in,height=2.0104in]{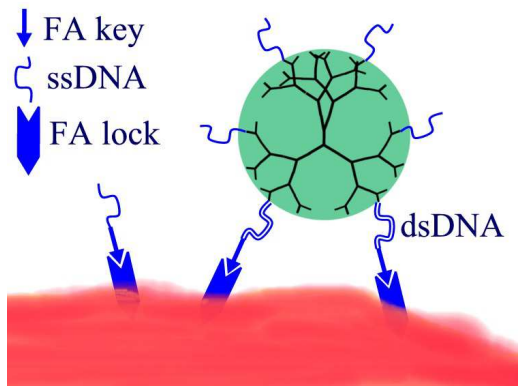}
\caption{(Color online). \ Single-stranded DNA (ssDNA) on the dendrimer
hybridize to the ssDNA attached to the folic acid (FA) key. \ }
\label{dendrimerscheme}
\end{figure}

The current experimental scheme uses direct targeting with folic acid ($%
\Delta _{FA}\simeq 17.5$), which does not optimize the coverage on cancerous
cells. \ By decreasing $\Delta $ the drug delivery can be tuned to the
favorable regime. \ To do so, consider binding to the cell through an
intermediary, perhaps single-stranded DNA (ssDNA). \ Instead of folic acid,
attach many identical sequences of ssDNA to the dendrimer. \ Then, one also
constructs a folic acid-ssDNA complex with the ssDNA sequence complementary
to that of the ssDNA attached to the dendrimer. \ The folic acid will bind
very strongly to the folic acid receptors on the cell membrane, leaving the
unhybridized ssDNA as a receptor (see Fig. \ref{dendrimerscheme}). \
Effectively one has replaced $\Delta _{FA}$ with a new value $\Delta _{DNA}$
which can be tuned very precisely by controlling the length and sequence of
the DNA. \ Due to the large degree of overexpression\cite{ovarian}, this
change substantially increases the ratio of dendrimers on cancerous to
normal cells. \ As indicated in Fig. \ref{logratio}, with $r\simeq 10$ there
is a $5$ fold improvement over direct targeting with folic acid!

In this work we presented a theoretical study of a cell-specific, targeted
drug delivery system. \ A simple "key-lock" model was proposed to determine
the effective dissociation rate and association rate constants of the
dendrimers as a function of the average number of folic acids, which permits
a direct comparison to the experimental results. \ The equilibrium coverage
of dendrimers on the cell surface was calculated, and the differences
between in vitro experiments and in vivo studies were discussed. \ The
degree of cooperativity of the drug delivery system is kinetically limited.
\ We quantified the notion of preferential selection of dendrimers to
cancerous cells, and demonstrated that the selectivity can be enhanced by
decreasing the strength of individual bonds. \ A particular implementation
of this idea using ssDNA was discussed. \ 

\begin{acknowledgments}
This work was supported by the ACS Petroleum Research Fund (PRF\ Grant No.
44181-AC10). \ We acknowledge B. Orr, M. Holl, P. Leroueil, and C. Kelly for
valuable discussions. \ 
\end{acknowledgments}

\bibliographystyle{apsrev}
\bibliography{acompat,dna}

\end{document}